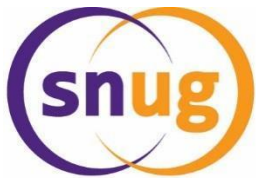

# Advanced In-Design Auto-Fixing Flow for Cell Abutment Pattern Matching Weakpoints

Yongfu Li, Valerio Perez, I-Lun Tseng,
Zhao Chuan Lee, Vikas Tripathi, Jason Khaw
and Yoong Seang Jonathan Ong

GLOBALFOUNDRIES
Singapore

**ABSTRACT**

Pattern matching design verification has gained noticeable attention in semiconductor technologies as it can precisely identify more localized problematic areas (weakpoints) in the layout. To address these weakpoints, engineers adopt "Rip-up and Reroute" methodology to reroute the nets and avoid these weakpoints. However, the technique is unable to address weakpoints due to the cell placement. The only present approach is to manually shift or flip the standard cells to eradicate the weakpoint. To overcome the challenge in going from a manual and laborious process to a fully automated fixing, we have proposed an in-design auto-fixing feature, tested with the commercial design tool, Synopsys IC Compiler. Our experimental result has demonstrated close to one hundred percent lithography weakpoints fixing on all of our 14nm designs.

# Table of Contents

# Table of Figures

# 1   Introduction

With the continuous scaling of the semiconductor technology, we are now facing the bottleneck of using 193i optical lithography process. Pattern-related defects continue to increase and limit the number of good die per wafer. The classical rule-based Design Rule Check (DRC) approach is no longer sufficient to guarantee 100% pattern printability. Therefore, starting from the GLOBALFOUNDRIES 40-nm technology, Design-for-Manufacturability (DFM) verifications, such as lithographic process variability simulation, chemical mechanical polishing (CMP) simulation, and critical area analysis (CAA) simulation, are required to identify manufacturing weak-points and prevent catastrophic errors such as open (necking) and shorts (bridging) issues, thus enabling early ramp to good yield [1], [2]. These DFM techniques provide many opportunities to improve layout where layout modifications can have a positive impact on the overall yield.

## 2 Present challenge & Propose advanced in-design auto-fixing flow for pattern matching weakpoints.

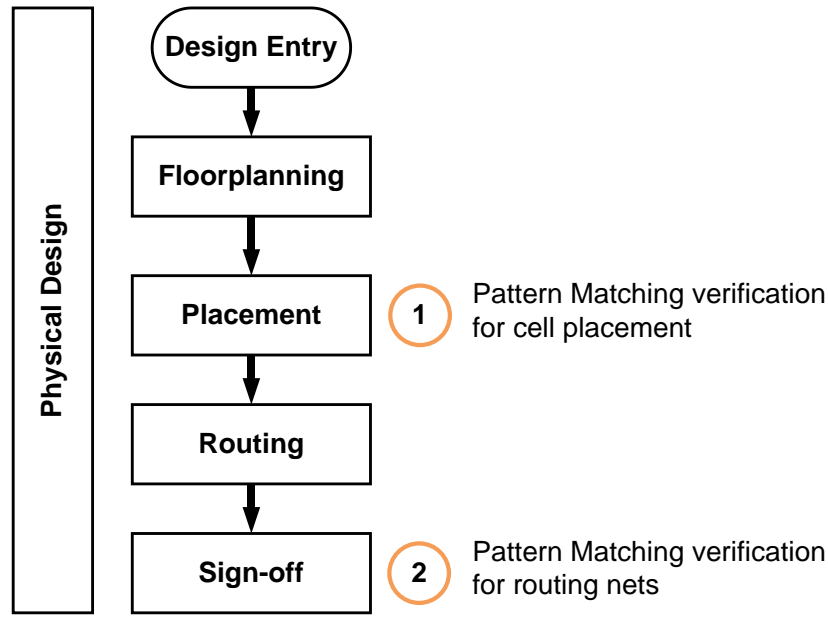

Figure 1. Workflow illustrating the use of pattern matching verification during design phase.

In a chip design, the physical design flow of an integrated circuit design is composed of the five main steps, namely (1) library preparation, (2) design floor-planning, (3) standard cell placement, (4) wire routing and (5) design sign-off (Figure 1). Physical verification is performed after physical design stage(post-layout) to assess the design manufacturability. In particular, pattern matching verification identifies layout patterns that are either difficult to print during lithography, or very susceptible to process variations. These patterns are commonly referred to as lithography weakpoint and it is required to identify and fix them early before the design goes into the production phase. In addition, pattern matching is preferred over DRC as it can identify more localized problematic areas in the layout, thus requires minimum changes to the layout [3].

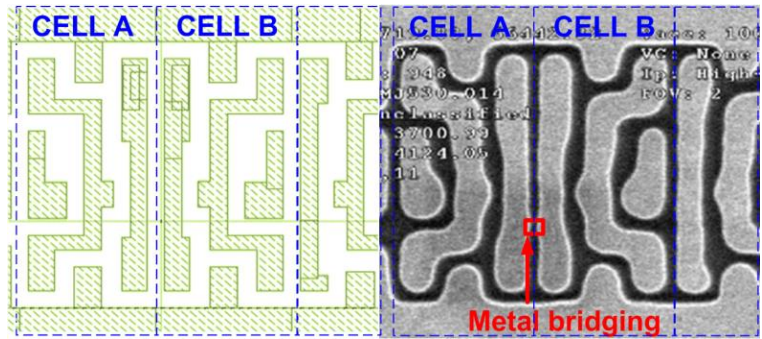

Figure 2. Layout and silicon image showing possible metal bridging weakpoint occurred at the cell abutment boundary.

To fix all lithography weakpoints in the design, engineers have to change the layout context at each lithography location. For example, a weakpoint located on the wire routing layer can be resolved with the "rip-up and reroute" methodology. However, when a weakpoint occurs at the boundary of the standard cells, as shown in Figure 2, the weakpoint has to be manually fixed by engineers through the shifting or flipping of the standard cells, followed by incremental routing to reconnect the wires to the pins. As engineers are always under enormous pressure to complete tasks, an automated fixing solution is required. To identify all lithography weakpoints, we recommend engineers to perform pattern matching verifications during the placement stage (marked in '(1)') and the sign-off stage (marked in '(2)'), as illustrated in Figure 1. Therefore, our proposed auto-fixing software can resolve all lithography, which occur at the boundary of the standard cells, to minimize performance impact on the circuit design,

## 3    Definitions of Cell's Placement and Orientation

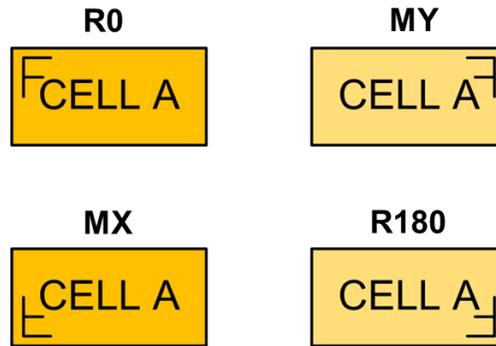

Figure 3. Four possible legal orientations of standard cell placement in the chip design

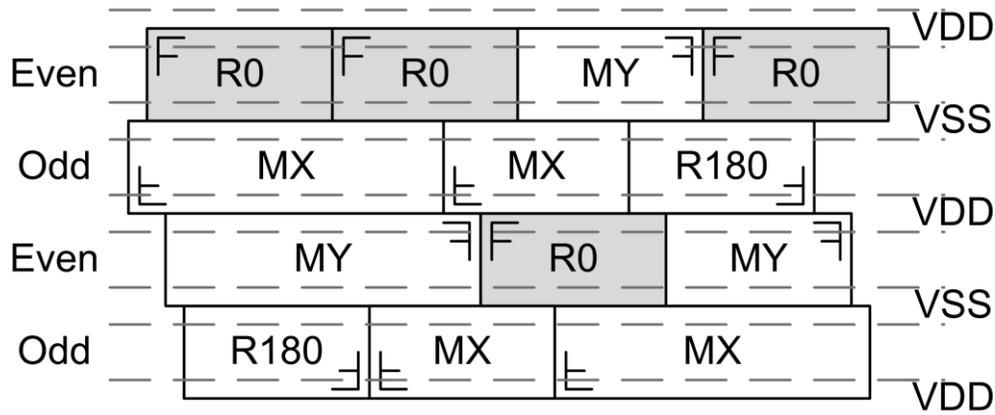

Figure 4. Standard cells placement in the chip design

In the chip design, there are four possible orientations of a standard cell that can be placed in the layout (Figure 3).  The R0- and R180-orientated cells are the default cell orientation and cell rotated by 180 degrees. The MX- and MY-orientated cells are cells mirrored along the X-axis and Y-axis, respectively. A conventional single-height standard cell is designed with the power signal (VDD) and the ground signal (VSS) at the top and bottom rails, respectively. Therefore, the single-height standard cells with R0- and MY-orientations are legally placed in the same row while the single-height standard cells with MX- and R180-orientations are placed at an alternate row. This arrangement allows the standard cells to be placed along pre-defined, interleaved horizontal VDD and VSS rails during the placement phase.

As shown in Figure 4, there is a space constraint to shift standard cell B to resolve the weakpoint (marked in '(a)'). As a result, the orientation of cell B has to be changed from MX- to R180-orientations in order to modify the layout context at the location.

# 4   Auto-Fixing Solution and Implementation

The associated IC compiler commands to implement the flow are summarized as follow and it can be broken down into the following phases [3].

A.   Perform physical verification (Pattern Matching) through  ICV
B.   Retrieve the errors from the error cell view
C.   Perform standard cell abutting fixing
D.   Re-iterate (A) –(C) to ensure no pattern matching errors.

```
# Setup ICV setting
set_physical_signoff_options \
    -default \
    -exec_cmd {icv} \
    -drc_runset "runset.rs" \
    -mapfile "map.file";

report_physical_signoff_options; # Report ICV setting

signoff_drc \
    -show_stream_error_environment false \
    -read_cel_view \
    -ignore_blockages_in_cells false \
    -run_dir "./result";
  set view [gui_open_error_view -name ${design}.err ];
  set types [list_drc_error_types -error_view $view ];

# Perform fixing
set error 1;
while {$error > 0} {
  set error 0;
  foreach type $types {
    set errors [get_drc_errors -type $type -error_view $view -quiet];
    foreach_in_collection id $errors {
      fix_cell_abutment -coordinates [get_attribute $id bbox];
      incr error;
    }
  }
 signoff_drc \
    -show_stream_error_environment false \
    -read_cel_view \
    -ignore_blockages_in_cells false \
    -run_dir "./result";
  set view [gui_open_error_view -name ${design}.err ];
  set types [list_drc_error_types -error_view $view ];
}
```

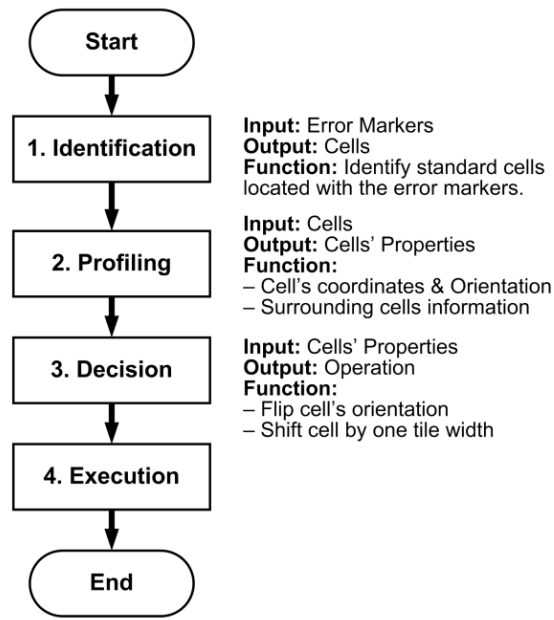

Figure 5. Workflow of our utility tool in the IC Compiler environment.

The proposed in-design auto-fixing algorithm is composed of four subroutines (Figure 5), which are listed as follows:

1. Identification
2. Profiling
3. Decision
4. Execution

As shown in Figure 4, the location of the weakpoint (error marker) can reside either at the boundary between the two affected cells (marked in '(a)') or within the affected cell (marked in '(b)'). The first subroutine, "Identification", single out the affected standard cells from the design based on the weakpoint's coordinates. When the boundary is located within the affected standard cell (marked in '(b)'), the subroutine extends the boundary and identifies the second affected standard cell. The second subroutine, "Profiling", obtains the standard cells' information such as boundary coordinates, orientation and understanding the available spaces adjacent to the affected standard cells. If there is no space adjacent to the affected standard cells, we can only resolve the weakpoint through cell flipping operation. Since there are three types of cell flipping operations, i.e. flip left and/or right cells, the subroutine, "Decision", which is based on "random walk" process [3], randomly determine one of the available operations for the next subroutine, "Execution", to execute. When the in-design auto-fixing algorithm is executed during the sign-off phase, engineers have to perform incremental routing to reconnect the wires to the pins.

# 5    Result & Discussion

The experiment has been carried out on a 14-nm standard cell library. We used our proprietary circuit generator software to design twenty circuits with netlist sizes ranging from 2,504 to 14,143,243 logic gates, as shown in Figure 6(a). The minimum and maximum initial violation counts recorded ranging from 2,504 to 193,076. As shown in Figure 6(b), the standard cell's placement utilization rate ranged from 65% to 85%.

We implemented our proposed algorithm and tested out on the commercial design tool, IC Compiler [4]. Figure 7 illustrates each of the intermediate steps to resolve a lithography weakpoint reported by the pattern matching verification tool. The weakpoint's location is situated at the boundary of two "NOR" logic gate (Figure 7 (a)). The first random operation is "flip the left cell" on mult_x_41/U525 (Figure 7 (b)) but due to the symmetrical layout of the logic gate, the weakpoint pattern context remains at the same location after the first iteration. The next iteration is "shift the left cell" operation on mult_x_41/U525 by one tile width (Figure 7 (c)). The space between the standard cells changes the layout context at the location which resolved the weakpoint.

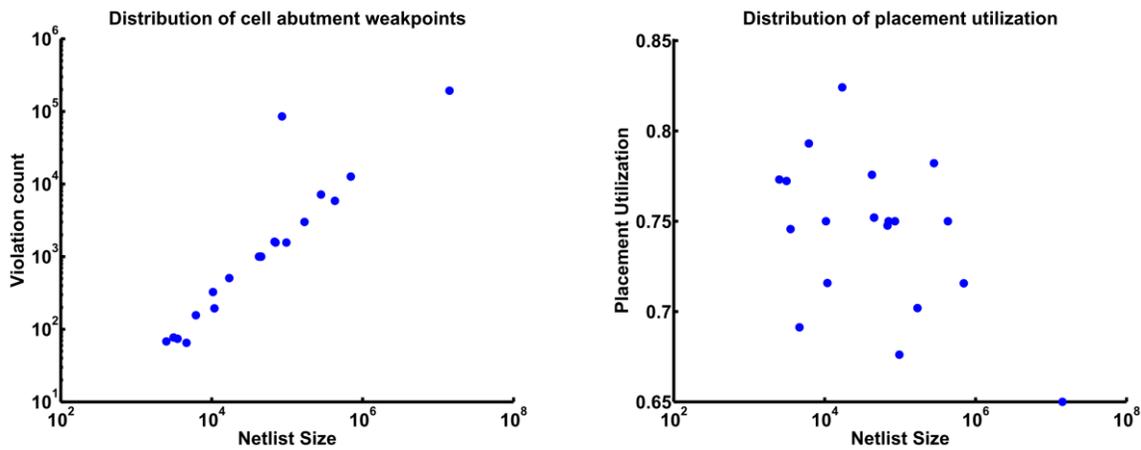

Figure 6. Distributions of netlist sizes against (a) the initial violation counts reported by pattern matching library (b) the standard cell placement utilization rate.

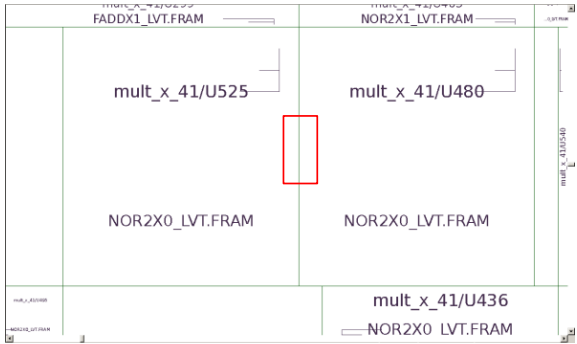

Coordinate: Error markers 🔲
    llx lly urx ury
Options: cell operation
    Flip Left cell
    Flip Right cell
    Flip Both cells
    Shifting cell

(a)

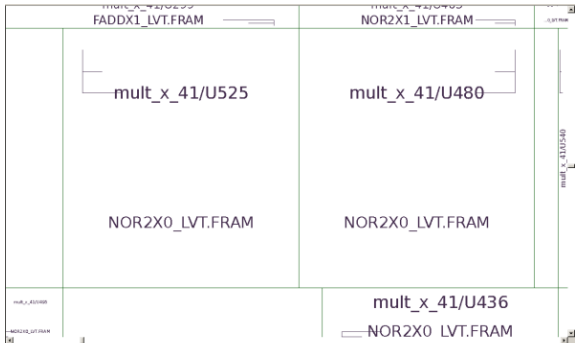

```
#-------------------------------------------------#
# Opertion: Flip left cell
#-------------------------------------------------#
# Cell Information/Attributes [Original]
# Cell:        mult_x_41/U525
# Bbox:        {77.976 168.872} {79.496 170.544}
# Orientation: FS
#-------------------------------------------------#
#-------------------------------------------------#
# Cell Information/Attributes [Modified]
# Cell:        mult_x_41/U525
# Bbox:        {77.976 168.872} {79.496 170.544}
# Orientation: S
#-------------------------------------------------#
```

(b)

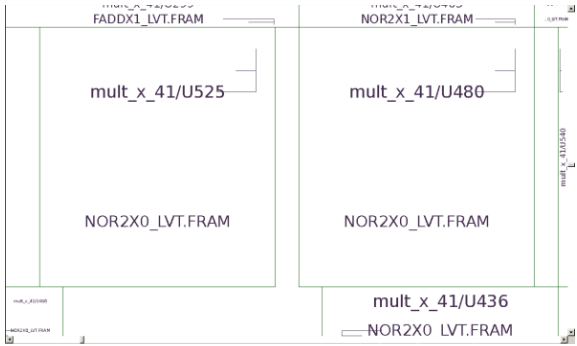

```
#-------------------------------------------------#
# Opertion: Shift the left cell mult_x_41/U525
#-------------------------------------------------#
# Cell Information/Attributes [Original]
# Cell:        mult_x_41/U525
# Bbox:        {77.976 168.872} {79.496 170.544}
# Orientation: S
#-------------------------------------------------#
#-------------------------------------------------#
# Cell Information/Attributes [Modified]
# Cell:        mult_x_41/U525
# Bbox:        {77.824 168.872} {79.344 170.544}
# Orientation: S
#-------------------------------------------------#
```

(c)

Figure 7. Layout view of the "NOR" logic gates (mult_x_41/U525 and mult_x_41/U480) (a) Location of the error marker between the logic gates (b) "flip left cell" operation on mult_x_41/U525 (c) "shift left cell" operation on mult_x_41/U525 by 1 tile width.

Since our software is based on "random walk" process, it is important to understand the effectiveness to resolve all the lithography weakpoints [3]. As shown in Figure 8, the software can achieve close to one hundred percent auto-fixing within four iterations except for the case where the number of weakpoints exceeded 193,706.

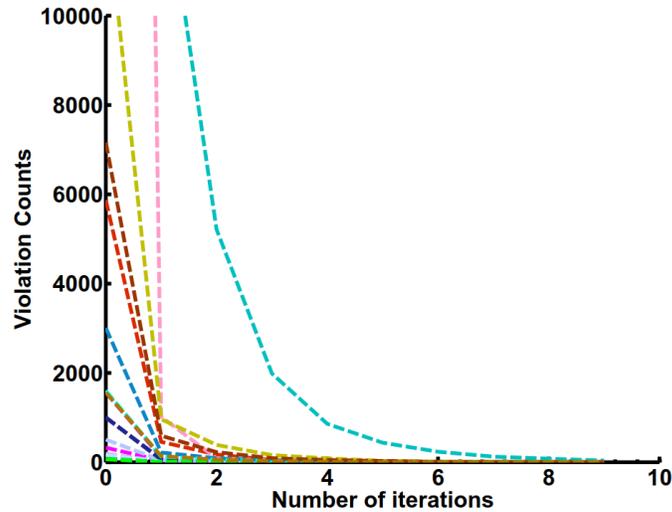

Figure 8. A Plot illustrating the number of weakpoints after each auto-fixing iteration.

## 6    Conclusions

In this work, we have proposed a new recommendation for pattern matching verification and have proposed an in-design auto-fixing flow to overcome the challenges in going from a manual and laborious process to a fully automated fixing. Our solution is implemented by using a commercial design tool, Synopsys IC Compiler, and our experimental result has demonstrated close to one hundred percent lithography weakpoints fixing on all of our 14nm benchmark designs.